# Integrated simulation of magnetic-field-assist fast ignition laser fusion

<u>T. Johzaki</u>[1], H. Nagatomo[2], A. Sunahara[3], Y. Sentoku[4], H. Sakagami[5], M. Hata[2], T. Taguchi[6], K. Mima[7], Y. Kai[1], D. Ajimi[1], T. Ishoda[1], T. Endo[1], A. Yogo[2], Y. Arikawa[2], S. Fujioka[2], H. Shiraga[2], H. Azechi[2], FIREX project

[1]Department of Mechanical Systems Engineering, Hiroshima University, Higashi-Hiroshima, Japan

[2]Institute of Laser Engineering, Osaka University, Suita, Japan

[3]Institute for Laser Technology, Suita, Japan

[4]Department of Physics University of Nevada Reno, Reno, NV, USA

[5]National Institute for Fusion Science, Toki, Japan

[6]Department of Electrical and Electronics Engineering, Setsunan University, Neyagawa, Japan

[7]The Graduate School for the Creation of New Photonics Industries, Hamamatsu, Japan

E-mail: tjohzaki@hiroshima-u.ac.jp

**Abstract**

To enhance the core heating efficiency in fast ignition laser fusion, the concept of relativistic electron beam guiding by external magnetic fields was evaluated by integrated simulations for FIREX class targets. For the cone-attached shell target case, the core heating performance is deteriorated by applying magnetic fields since the core is considerably deformed and the most of the fast electrons are reflected due to the magnetic mirror formed through the implosion. On the other hand, in the case of cone-attached solid ball target, the implosion is more stable under the kilo-tesla-class magnetic field. In addition, feasible magnetic field configuration is formed through the implosion. As the results, the core heating efficiency becomes double by magnetic guiding. The dependence of core heating properties on the heating pulse shot timing was also investigated for the solid ball target.

## 1. Introduction

In fast ignition scheme of inertial confinement fusion, laser-accelerated particle beams (electron and/or ion beams) heat a pre-compressed fusion fuel up to the ignition temperature. More than 10% of energy coupling efficiency of a heating laser to a compressed dense core plasma $\eta_{laser \to core}$ is essential for that the fast ignition prevails the conventional central ignition scheme. To demonstrate the efficient core heating, The Fast Ignition Realization Experiments project (FIREX project) is now being conducted using the Gekko XII (GXII) + LFEX laser system at the Institute of Laser Engineering, Osaka University [1]. The integrated simulations [2] and experimental data [3] showed very low heating efficiency ($\eta_{laser \to core} < 1$ %). There are two crucial issues preventing efficient heating; one is too hot fast electron temperature, and the other is too large beam divergence. By improving the contrast ratio of heating laser pulse, we have eliminated the plasma formation inside the guiding cone before the main part of the heating pulse comes, and successfully reduced the energy (or temperature) of fast electrons with keeping the energy convergence from laser to relativistic electron beam (REB) as constant [4]. As for the beam divergence, instead of the control of laser-plasma interactions to supress the angular spread of the fast electron distribution function, some guiding ideas for fast-electron beam with large beam divergence have been proposed, e.g., guiding schemes using self-generated magnetic fields (resistive guiding [5-11] and vacuum gap guiding [2, 12, 13]) and using the externally-applied magnetic fields [14-18]. In the present paper, the latter scheme is discussed.

The required magnetic field strength for REB guiding has been evaluated as a function of irradiating laser intensity by 2-dimensional (2D) collisionless particle-in-cell (PIC) simulations [17]. It was shown that the kilo-tesla-class magnetic field is needed for FIREX class experiments where the heating pulse intensity is of the order of $10^{19}$W/cm$^2$. The generation of kilo-tesla magnetic field and the electron beam guiding by applying external magnetic field have experimentally demonstrated using laser-driven capacitor-coil scheme [3, 19, 20]. Not only the strength of magnetic field but also the field configuration is important. The externally applied magnetic field will be compressed by fuel implosion except for the region inside the guiding cone, which results in forming a mirror field configuration. Under the mirror field configuration, we will expect the beam focusing while at the same time concerning about the mirror reflection. The effects of mirror configuration were also discussed on the basis of the 2D PIC simulations [18], which showed the sufficient guiding performance in collisional dense plasma by kilo-tesla-class external magnetic fields for the case with moderate mirror ratio ($\leq 10$). These numerical and experimental investigations showed the potential of beam guiding by external magnetic fields. For further study, we conducted integrated simulations for magnetic-field-assist fast ignition laser fusion. In the following, we discuss the core heating properties for the dense core with the converging magnetic field configuration appeared in the implosion of cone-attached targets at FIREX class experiments.

## 2. Simulation model

In the integrated simulations, first, we carried out the implosion simulations to evaluate the imploded core and magnetic field profiles around the maximum compression, where a radiation-hydro code PINOCO [21] was used. PINOCO is a single-fluid two-temperature hydrodynamics code written in 2D spherical geometry (*r-θ*) and includes the energy transports due to radiation (flux-limited multi-group diffusion model), thermal conduction (flux-limited spritzer-Harms model), and laser pulse (ray-trace model). To consider the magnetic field effects, it has been extended to be a resistive magnetohydrodynamic one [22]. We set the implosion parameters by assuming GXII laser system condition at ILE Osaka University, i.e., we assumed the laser energy of 1.5 kJ, the wavelength of 0.53 μm, Gaussian pulse with 1.3 ns FWHM and peak timing of t = 1.5 ns.

The following core heating simulations were carried out by using FIBMET [23] which is a hybrid-type code written in cylindrical geometry (*r-z*). The bulk plasma is treated by a fluid model and the transport of fast particles (such as a REB) is treated by particle scheme. The resistive field model is adopted. The collision between the fast particles and the bulk plasma is treated by the Fokker-Planck collision model. In the core heating simulations, we used the imploded core and magnetic field profiles obtained from the implosion simulations. The REB was injected at the tip of cone by assuming the experimentally observed profiles [4], where the energy spectrum $f(E) \propto 0.95\exp(-E/0.9) + 0.05\exp(-E/5)$ (*E* is electron energy in MeV). The beam divergence is assumed as 45 degree half angle. For the temporal and spatial profiles, we assumed the Gaussian with 1 ps duration (FWHM) and the super Gaussian with 20 μm radius (HWHM), respectively. The REB peak intensity and the injected energy are $1.55 \times 10^{19}$W/cm$^2$ and 410 J, respectively.

## 3. Simulation results

### *3.1.* **Spherical shell target**

The initial profile of a fuel target is shown in **Fig.1** (a). The CD shell radius and thickness are 250 μm and 8 μm. For this shell a gold cone with open angle of 45 degree (full angle) is attached. The cone tip with 7 μm thick and 40 μm diameter is located at 50 μm away from the shell center. The external magnetic field $B_{z,\text{ext}} = 200$ T is uniformly applied in *z*-direction. Figure 1 (b) shows the imploded core profile (upper) and magnetic field configuration (lower) at $t = 1.87$ ns. At this moment, the fuel does not reaches the maximum compression yet. Due to the suppression of thermal flow perpendicular to the magnetic field lines, the pressure unbalance is imposed, which induces the hydrodynamic instability. As the results, the equator part of the shell, which is the region perpendicular to *z*-axis in Fig.1, falls into the shell center faster than the other region and collapses at the center. Thus, the strongly deformed core is formed [22]. This asymmetric implosion leads to the strongly deformed the magnetic field configuration around the core center. In addition, the jet-like plasma flow generated due to the shell collapse pushes the magnetic field on the cone tip, and forms the magnetic field wall on the cone tip, where the magnetic field strength reaches ~ 20 kT, which is about 100 times higher than the initial value. On the other hand, the magnetic field strength inside the cone where the fast electrons are generated is kept as the initial value since the plasma does not move. Such a magnetic field configuration is not feasible to the fast electron beam guiding.

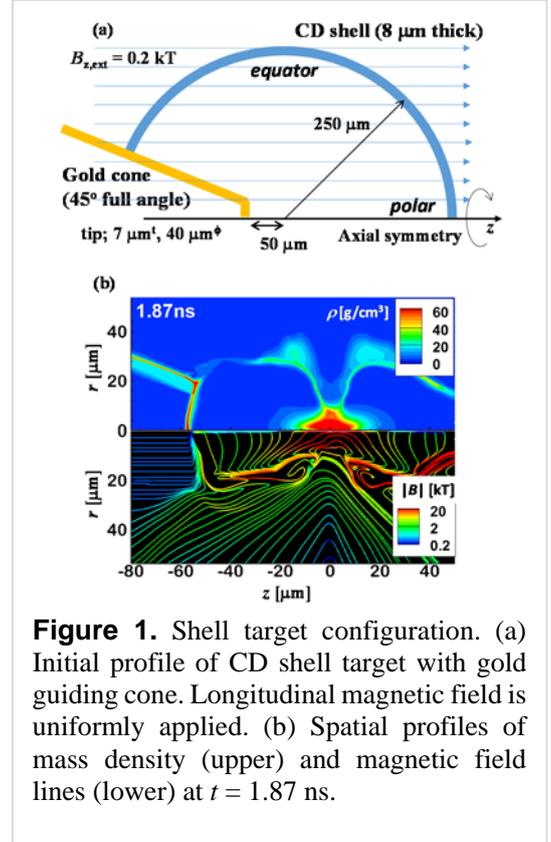

**Figure 1.** Shell target configuration. (a) Initial profile of CD shell target with gold guiding cone. Longitudinal magnetic field is uniformly applied. (b) Spatial profiles of mass density (upper) and magnetic field lines (lower) at $t = 1.87$ ns.

To evaluate the effects of external magnetic field on the core heating process, we carried out the REB transport simulations using the core plasma and magnetic field profiles. The REB was injected at the tip of cone with the profiles noted in Sec.2. **Figure 2** shows the spatial profiles of the fast electron energy density at $t_{\text{tran}} = 3$ ps, where $t_{\text{tran}}$ is time in the transport simulation and $t_{\text{tran}} = 0$ is the start of the transport simulation. If the external magnetic field is neglected in the transport simulation (Fig.2(a)), the fast electrons flow forward and spread throughout the simulation region. Because of the small core size and of the large beam divergence, the heating efficiency is small; the core heating efficiency, that is the ratio of the REB energy deposited in the dense core region ($\rho > 5$ g/cm$^3$) to the injected REB energy, is $\eta_{\text{REB}\rightarrow\text{core}} = 1.9$ %. In the case of the simulation with the external magnetic field, since the magnetic field strength around the REB injected region (the cone tip), $B_{z,\text{inj}} = 0.2$ kT, is too small to trap MeV electrons, we artificially intensified the magnetic field strength; the magnetic field strength was multiplied tenfold in whole region to make $B_{z,\text{inj}} = 2$ kT. The result of the simulation showed, as was predicted, that the most of fast electrons cannot overcome the magnetic field wall ahead of the cone tip and are reflected. As the result, the heating efficiency becomes much lower than that for the case without the external field, $\eta_{\text{REB}\rightarrow\text{core}} = 0.0015$ %.

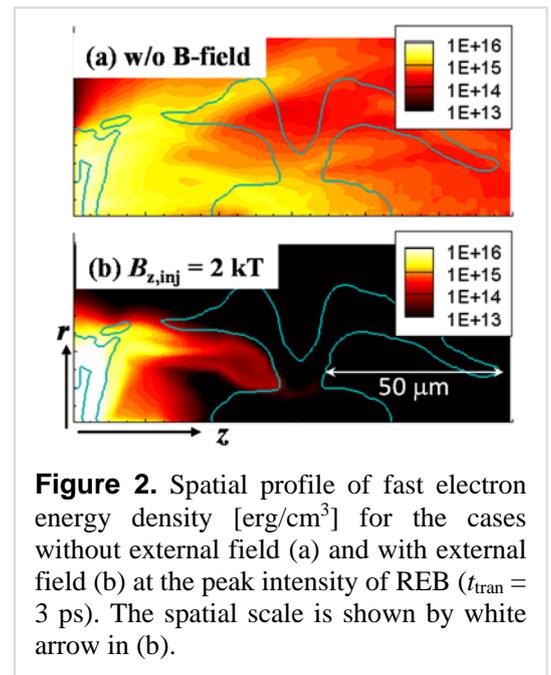

**Figure 2.** Spatial profile of fast electron energy density [erg/cm$^3$] for the cases without external field (a) and with external field (b) at the peak intensity of REB ($t_{\text{tran}} = 3$ ps). The spatial scale is shown by white arrow in (b).

The above simulation results showed that the shell implosion case, application of external field not only disturbs the high density core formation but also prevents the efficient core heating. To apply the magnetic field to the shell target, the applying magnetic field structure and applying timing should be optimized. Otherwise, the external field negatively affects the implosion and core heating profiles.

## 3.2. Spherical solid target

A shell target is used in the conventional central spark ignition scheme to form the central hot spot and surrounding cold dense main fuel structure through the implosion. In the fast ignition scheme, since the implosion is needed only to form a high-density fuel core, but not to generate a hot spot for initiation of fusion burning, an alternative implosion scheme is applicable. To generate the high-density core stably more than a shell implosion, we have proposed use of a spherical solid target [3] where a fuel target is compressed to high density by a spherically-converging shock. The formation of dense core using a simple solid ball target [3] and a solid target with a gold cone [24] has been experimentally demonstrated. In this section, we evaluate the implosion and core heating properties for a solid ball target with external magnetic fields.

### 3.2.1. Implosion simulations

The initial target configuration is shown in **Fig.3**. The target is a spherical CD solid ball with 100 μm of radius. A gold cone is embedded in the CD target to reduce the distance between the cone tip and the target center and also to prevent the cone tip from breaking due to the attack of converging shock. To protect the cone wall against the ablated plasma, the radiation from the coronal plasma and the reflected laser light, the surface of cone wall is coated by the CD layer with 20 μm thick. For this target, the external magnetic field is uniformly applied in z-direction with the strength of $B_{z,ext} = 0 \sim 2$ kT. The imploded core profiles at the maximum compression are shown for $B_{z,ext} =$ 0, 0.5 1.0, 2.0 kT in **Fig.4**. Here the maximum compression

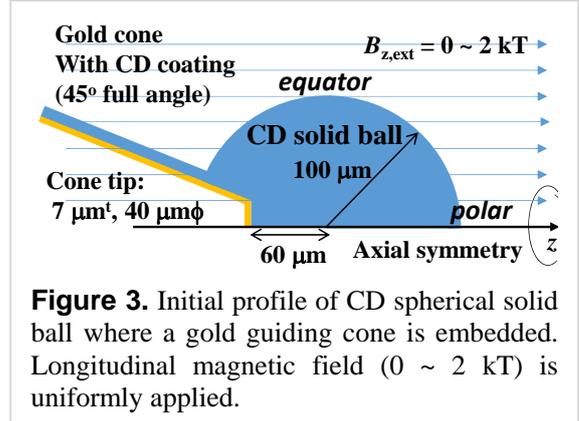

**Figure 3.** Initial profile of CD spherical solid ball where a gold guiding cone is embedded. Longitudinal magnetic field (0 ~ 2 kT) is uniformly applied.

is defined as the timing when the integration of mas density $\rho(r, z)$ for the fuel region along z-axis at $r = 0$, $\int_{CD} \rho(r=0, z)dz$, becomes maximum. Compared with the shell target case, the effect of external field is small for the solid target case. As is the case with a shell implosion, however, since the ablation pressure becomes higher for the equator region with increasing the external field strength, so the incoming shock is stronger for the equator region. As the result, with increasing $B_{z,ext}$, the timing of maximum compression $t_{max}$ becomes earlier ($t_{max}$ =1.88 ns for $B_{z,ext} = 0$ to $t_{max}$ =1.64 ns for $B_{z,ext} = 2$ kT), the maximum density $\rho_{max}$ becomes larger

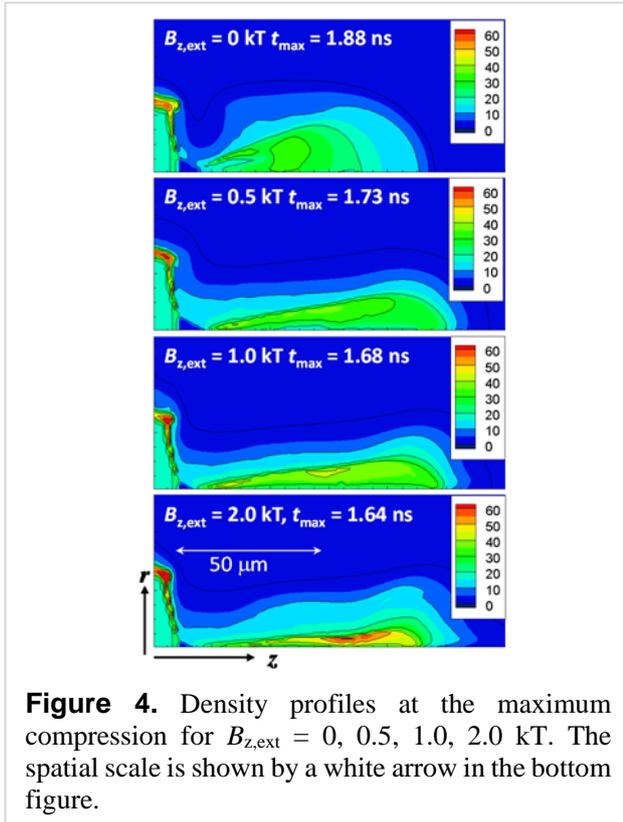

**Figure 4.** Density profiles at the maximum compression for $B_{z,ext} =$ 0, 0.5, 1.0, 2.0 kT. The spatial scale is shown by a white arrow in the bottom figure.

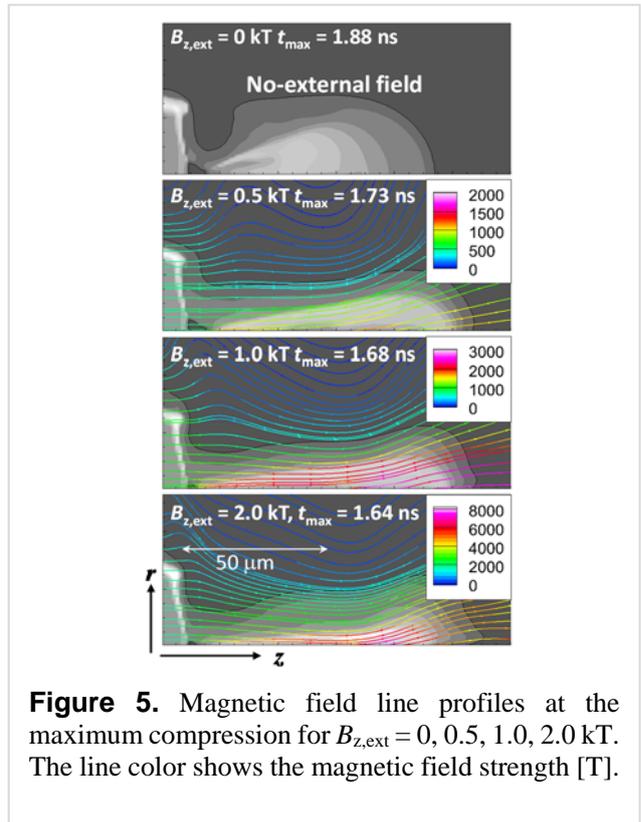

**Figure 5.** Magnetic field line profiles at the maximum compression for $B_{z,ext} =$ 0, 0.5, 1.0, 2.0 kT. The line color shows the magnetic field strength [T].

($\rho_{max}$ = 34 g/cm$^3$ for $B_{z,ext}$ = 0 to $\rho_{max}$ = 61 g/cm$^3$ for $B_{z,ext}$ = 2 kT), and the core shape becomes more elongated one. The magnetic field line structures at the same timing are shown in **Fig.5**. Compared with the shell implosion case, the magnetic field lines are much smoother and the mirror ratio (the ratio of magnetic field strength between the cone tip region and the dense core region) is smaller. The mirror ratio is 2.4, 2.9 and 3.8 for $B_{z,ext}$ = 0.5, 1.0 and 2.0 kT, respectively. Compared to the density compression ratio, the compression of magnetic field is small for all cases. As was discussed in Ref. 3, this is due to the small magnetic Reynolds number (close to unity), and then the compression of magnetic field is not efficient compared with the ideal magnetohydrodynamics case. The smooth magnetic field line profile and the moderate mirror ratio obtained for the solid ball target are adequate for the REB guiding.

### 3.2.2. Core heating properties

Using the profiles of imploded core and magnetic field at the maximum compression obtained from implosion simulations, we carried out the REB transport simulations to evaluate the guiding effects for the solid target case. The injected REB condition was described in Sec.2. **Figure 6** shows the spatial profiles of fast electron energy density at the peak intensity of REB ($t_{tran}$ = 3 ps). In the case without external field, the REB spatially diverges with its propagation due to the large angular spread. Around the central axis, there is the region where the beam energy density becomes low. This is due to the scattering of fast electrons by the magnetic field generated around the core by the Biermann battery effect. Due to the REB heating, the direction of bulk electron temperature gradient $\nabla T_e$ becomes non-parallel to that of bulk electron density gradient $\nabla n_e$ around the dense core edge [23], which causes the Biermann battery fields in the direction to scatter the fast electrons. For the case of $B_{z,ext}$ = 0.5 kT, the guiding effect can be observed. The guiding by the external fields weakens the scattering effect due to the Biermann battery field. However, the magnetic field strength at the REB injection region ($B_{z,ext}$ = 0.5 kT) is not strong enough to trap the MeV-class fast electrons. Besides the core shape is elongated by the external field. Hence the pronounced enhancement of the core heating efficiency by applying external field is not observed. For the case of $B_{z,ext}$ = 1.0 kT, the guiding effect by external field overcomes the scattering effect by the Biermann battery field. Although the fast electrons in the beam edge region (large $r$ region) are not guided to the core, more fast electrons are guided to the core compared to the case of $B_{z,ext}$ = 0.5 kT, and then the heating efficiency becomes clearly higher than that for the case without external field. For further higher $B_{z,ext}$ case, the guiding becomes more remarkable.

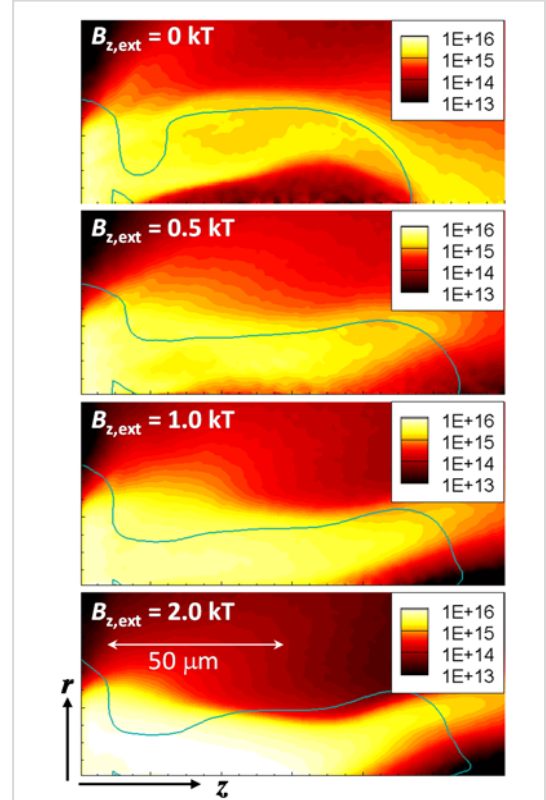

**Figure 6.** Spatial profiles of fast electron energy density [erg/cm$^3$] at the peak intensity of REB ($t_{tran}$ = 3 ps) for $B_{z,ext}$ = 0, 0.5, 1.0, 2.0 kT.

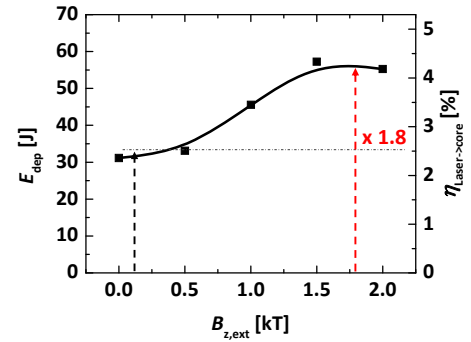

**Figure 7.** Energy deposited by REB in the dense core region ($\rho$ > 5 g/cm$^3$) $E_{dep}$ as a function of $B_{z,ext}$. The right axis shows the energy coupling efficiency of the heating laser to the dense core by using experimentally measured energy conversion efficiency of heating laser to the REB, 31% [3].

In **Fig.7**, the energy deposited by the REB in the dense core ($\rho$ > 5 g/cm$^3$) $E_{dep}$ is plotted as a function of $B_{z,ext}$. The right axis shows $\eta_{laser \rightarrow core}$ which is the energy coupling efficiency from the heating laser to the dense core by using experimentally measured energy conversion efficiency of heating laser to the REB, $\eta_{laser \rightarrow REB}$ = 31% [25]. As was noted above, in the case of a solid ball CD target, we can enhance the heating efficiency by applying the longitudinal external field. To obtain clear enhancement, kilo-tesla-class field is required, *e.g.*, the application of 1.5 kT external field doubles the heating efficiency roughly (from $\eta_{laser \rightarrow core}$ = 2.3 % for $B_{z,ext}$ = 0 kT to $\eta_{laser \rightarrow core}$ = 4.3 % for $B_{z,ext}$ = 1.5 kT) . This requirement for external field is consistent with the evaluation of guiding performance by 2D PIC simulations [17, 18].

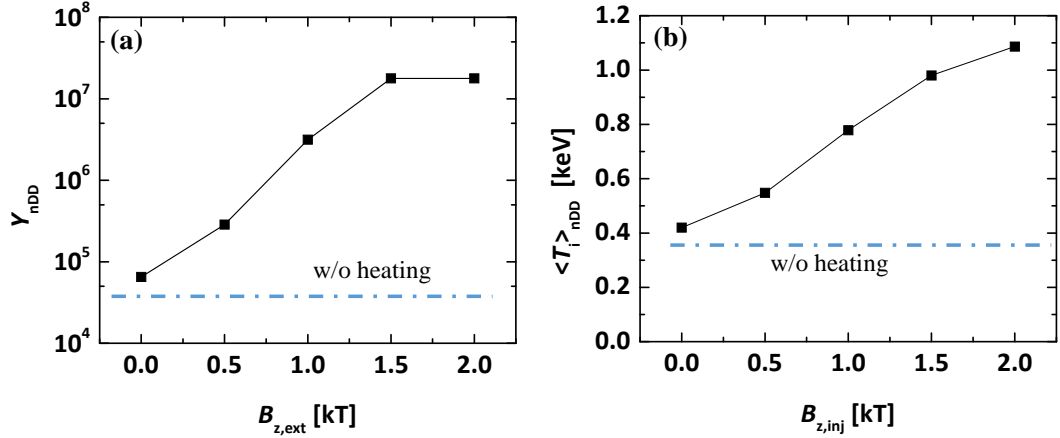

**Figure 8.** Neutron yield from D(d,$^3$He)n reactions $Y_{nDD}$ and neutron-weighted average ion temperature $<T_i>_{nDD}$ as a function of $B_{z,ext}$. The values for the case without REB heating is shown by dot-and-dash lines.

The neutron yields from D(d,$^3$He)n reactions $Y_{nDD}$ and the neutron-averaged ion temperature $<T_i>_{nDD}$ obtained from the core heating simulations are plotted as a function of $B_{z,ext}$ in **Fig.8**. The definition of $<T_i>_{nDD}$ is

$$<T_i>_{nDD} = \frac{\iint T_i(r,z,t) R_{nDD}(r,z,t) dV dt}{\iint R_{nDD}(r,z,t) dV dt},$$

where $R_{ndd}(r, z, t)$ is the D(d,3He)n fusion reaction rate. The values for the case without REB heating is indicated by dot-and-dash lines. For the case without REB heating, $Y_{nDD}$ is less than $10^5$ and $<T_i>_{nDD}$ is less than 0.4 keV. When the REB is injected, the both values are increased. But, for the case without external field case, the enhancements are very low. With increasing the applying field strength, the enhancements become remarkable. When the applied magnetic field strength is $B_{z,ext} = 2$ kT, $Y_{nDD} \sim 10^7$ and $<T_i>_{nDD} = 1.1$ keV are obtained; $Y_{nDD}$ is enhanced by more than 2 order of magnitude and enhancement of $<T_i>_{nDD}$ is about 0.7 keV.

### 3.2.3. Heating pulse shot timing

In the fast ignition scheme, the heating pulse shot timing is important to heat the fuel and initiate the fusion burning. Unfortunately, the energy of implosion and heating lasers used at the FIREX-class experiments is not large enough to initiate the fusion burning. In such a case, the margin of the heating shot timing may be limited by the lifetime of the dense core. In a practical case, there is uncertainty in control of shot timing. So the sensitivity evaluation is important for the practical purpose. Also, it is interesting how the shot timing affects the hydrodynamics of the core. So we carried out the REB transport simulations by changing the REB injection timing. **Figure 9** shows the temporal evolution of the imploded core for $B_{z,ext} = 1.0$ kT. In Fig.9(a), $\rho R_{z,max}$ and $\rho R_{r,max}$ are defined as

$$\rho R_{z,max} = \max[\rho R_z(r)] = \frac{1}{2}\max\left[\int_{CD}\rho(r,z)dz\right],$$

$$\rho R_{r,max} = \max[\rho R_r(z)] = \max\left[\int_{CD}\rho(r,z)dr\right],$$

where $\int_{CD}\rho(r,z)dz$ and $\int_{CD}\rho(r,z)dr$ means the integral over the CD plasma region. The maximum compression is achieved at $t$ = 1.68 ns. Before this timing, the incoming shock does not reach the center. After the maximum compression ($t$ = 1.73 ns), the fuel disassembly already started.

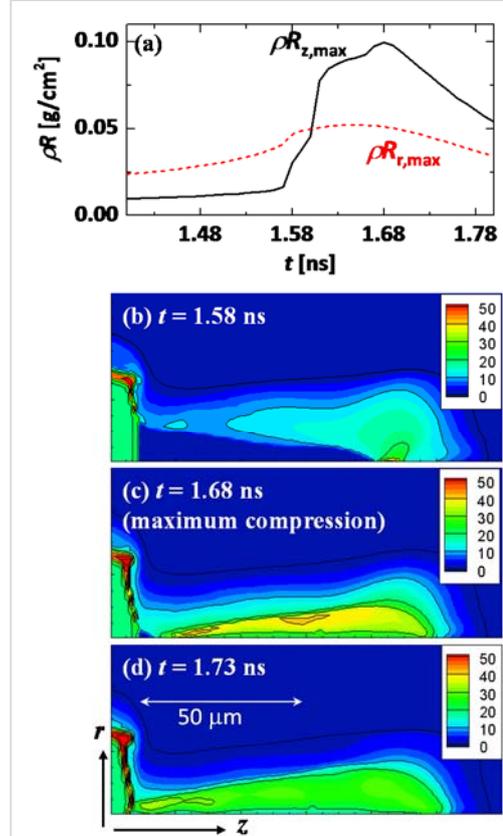

**Figure 9.** Temporal evolution of an imploded core for $B_{zext} = 1.0$ kT. Fuel rR vs. time (a), Spatial profiles of mass density [g/cm$^3$] at $t = 1.58$ ns (b), 1.68 ns (c) and 1.73 ns.

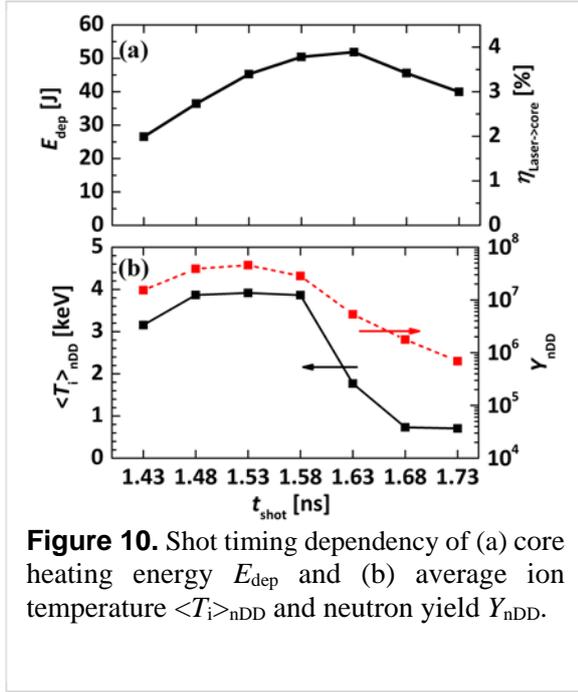

**Figure 10.** Shot timing dependency of (a) core heating energy $E_{dep}$ and (b) average ion temperature $<T_i>_{nDD}$ and neutron yield $Y_{nDD}$.

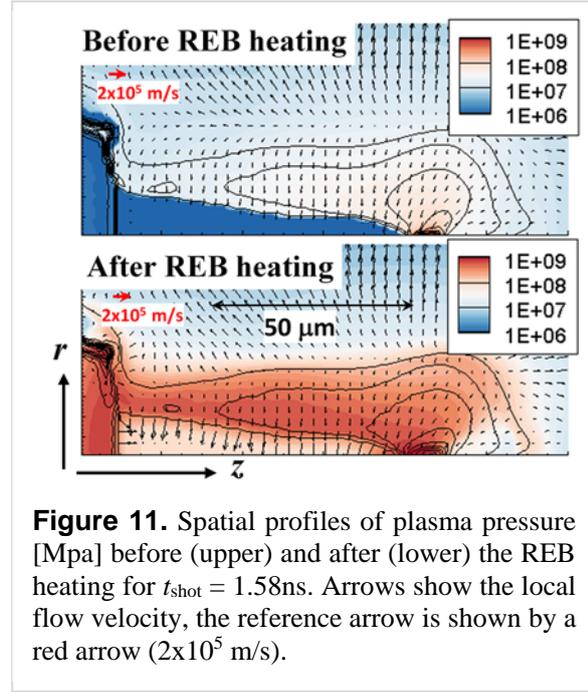

**Figure 11.** Spatial profiles of plasma pressure [Mpa] before (upper) and after (lower) the REB heating for $t_{shot}$ = 1.58ns. Arrows show the local flow velocity, the reference arrow is shown by a red arrow ($2 \times 10^5$ m/s).

The REB transport simulations were carried out for the timing from $t$ = 1.43 ns (250 ps before the maximum compression) to $t$ = 1.73 ns (50 ps after the maximum compression), every 50 ps. The core heating energy $E_{dep}$, neutron yields $Y_{nDD}$ and averaged ion temperature $<T_i>_{nDD}$ are plotted as a function of shot timing $t_{shot}$ in **Fig.10**. The core heating energy $E_{dep}$ becomes maximum when the REB is injected at 50 ps before the maximum compression ($t_{shot}$ = 1.63 ns) since the size of core in radial direction is larger at $t$ = 1.63 ns than that at $t$ = 1.68 ns in spite of small differences in $\rho R_{z,max}$ and $\rho R_{r,max}$. Though the coupling efficiency is not so sensitive to the shot timing (e.g., the shot timing window for $\eta_{laser \to core}$ > 3 % is about 200 ps, $t$ = 1.53 ns ~ 1.73 ns), the heating pulse should be shot around the maximum compression to maximize the core heating energy. On the other hand, the REB injection earlier than the maximum compression results in the higher $Y_{nDD}$ and $<T_i>_{nDD}$ compared to those obtained at the REB injection at the maximum compression. This is due to the intensification of incoming shock by the REB heating. The pressure profiles before (upper) and after (lower) the REB heating are shown in **Fig.11**. **Figure 12** show the temporal evolution of spatial profiles of ion temperature and fusion reaction rate. Due to the REB injection, the low density plasma in front of the incoming shock, where the density is kept as the initial value (solid density), is pre-heated. The REB heating of high dense plasma behind the shock is much higher than that in front of the shock, which intensifies the incoming shock. After the REB heating, thus, the intensified incoming shock is collapsed at the central axis, which hydro-dynamically heats the pre-heated low-density plasma. This shock collapse heating leads to neutron flash from the locally-heated region. In such a case, most of neutrons come from this locally-heated central region, and then the neutron-weighted average ion temperature is dominated by the temperature around there. This effect can be expected when the REB (or other heating pulse) is injected before the maximum compression. This heating scheme is the combination of direct REB heating and shock-collapse heating. Usually, the REB heats the bulk electron, and the bulk ion is heated by the collisional relaxation process. This relaxation time is not so small for the FIREX-class relatively low density plasma. Before the sufficient relaxation is achieved, the fuel disassembly will started, which results in the un sufficient ion heating. However, the present shock collapse heating does not require the temperature relaxation. The shock intensification is caused by the increase in the pressure of bulk electron heated by the REB, and the ion is hydrodynamically heated due to the collapse of the intensified incoming shock. So a high heating efficiency for bulk ion can be expected in this scheme.

## 4. Conclusion

For enhancing the core heating efficiency in relativistic fast electron beam (REB) driven fast ignition laser fusion, we have proposed to externally apply longitudinal magnetic field for REB guiding to the dense core. On the basis of integrated simulations, we evaluated the effects of application of external field on implosion and core heating processes for a cone-attached spherical CD shell and solid ball targets in FIREX-class experiments.

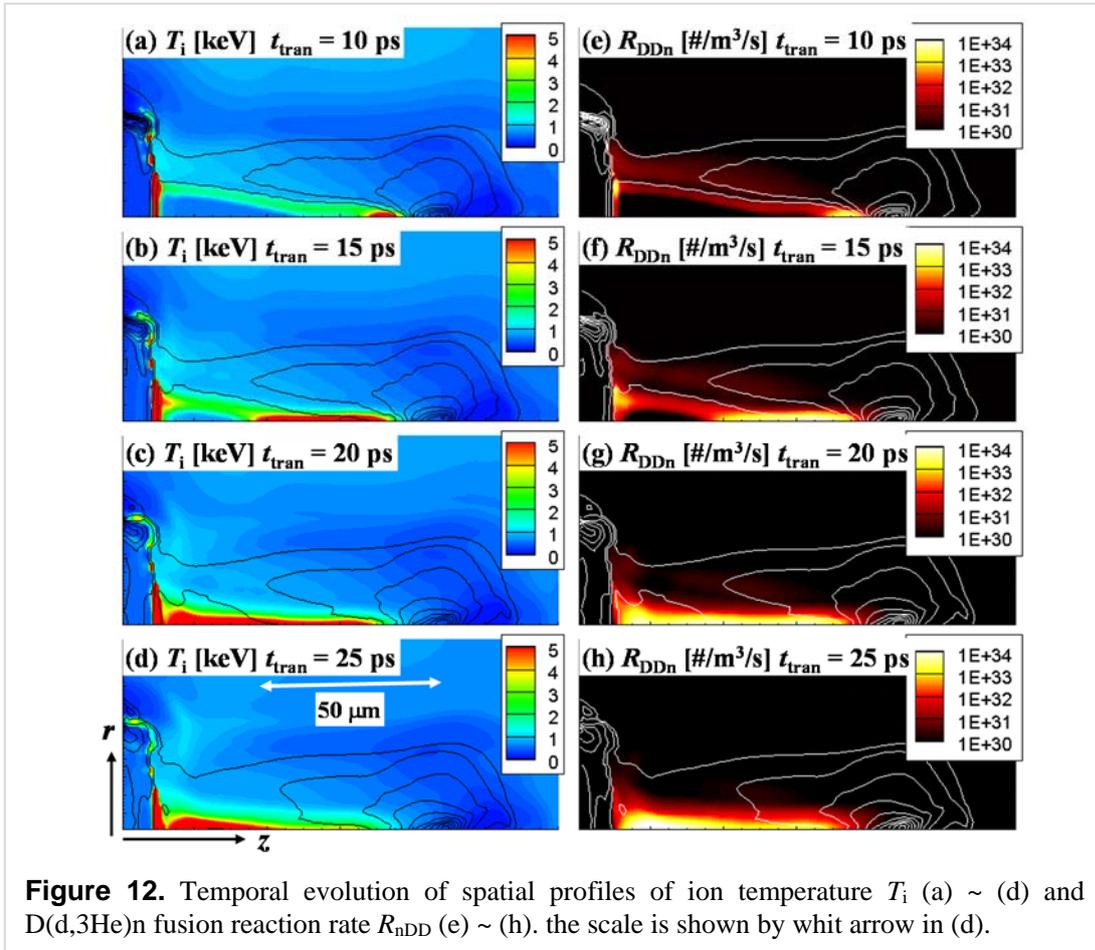

**Figure 12.** Temporal evolution of spatial profiles of ion temperature $T_i$ (a) ~ (d) and D(d,3He)n fusion reaction rate $R_{nDD}$ (e) ~ (h). the scale is shown by whit arrow in (d).

In the case of a shell target, it is found that the inhibition of thermal flow perpendicular to the magnetic field lines results in significant deformation of a fuel core. In addition, the magnetic wall is formed ahead of the cone tip, which reflects the almost all fast electrons. As the results, the application of magnetic field negatively affects the implosion dynamics and core heating properties.

On the other hand, for the solid ball target case, the implosion is more stable though the deformation due to the application of magnetic field occurs to a certain extent (*i.e*., the core shape is elongated). In addition, the magnetic field configuration feasible for REB guiding is formed through the implosion such that the magnetic field lines are so smooth and the mirror ratio is moderate ($R_m \sim 3$). The core heating simulations showed that application of kilo-tesla-class magnetic field enhances the core heating efficiency, and the kilo-tesla-class field is required to obtain the clear enhancement. Also, we evaluated the dependence of core heating properties on the heating pulse shot timing. In FIREX-class experiments, the heating pulse should be shot around the maximum compression to obtain the high coupling efficiency and the shot timing window is about 200 ps. The earlier shot timing than the maximum compression leads to higher neutron yield and neutron-average ion temperature since the incoming shock is intensified by REB heating.

The experimental demonstration will be carried out in the near future experiments at GekkoXII + LFEX laser systems, ILE Osaka University. The ignition and high gain design using solid ball target with external field is of intrinsic importance to drive the fast ignition research, and now we are engaged in the ignition design on the basis of integrated simulation.

**Acknowledgments**

This work is partially conducted under the joint research projects of the ILE, Osaka Univ. (FIREX-project), and with the supports of the NIFS Collaboration Research program (NIFS12KUGK057, NIFS15KUGK087, NIFS15KUGK093, NIFS15KUGK094, NIFS14KNSS054) and JSPS KAKENHI (25400539, 25400534, 26400532, 15H03758, 16H02245, 16K05638, 15K17798). We are grateful for the support of the computer room of ILE and the cybermedia center, Osaka University.